\documentclass[pra,twocolumn,superscriptaddress,showpacs,preprintnumbers,amsmath,amssymb]{revtex4-1}
\usepackage{graphicx}% Include figure files
\usepackage{dcolumn}% Alig table columns on decimal point
\usepackage{bm}% bold math
\usepackage{enumerate}
\usepackage{mathrsfs}
\usepackage{amsmath}
\usepackage{epstopdf}
\usepackage[titletoc]{appendix}
\usepackage[colorlinks=true,breaklinks=true,linkcolor=blue,citecolor=blue,urlcolor=blue]{hyperref}
\usepackage{color}
\usepackage{changes}
\newcommand{\Tr}{\rm Tr}

\begin{document}
%\bibliographystyle{apsrev} % Choose Phys. Rev. style for bibliography, Rev.4
%\preprint{APS/123-MW}
\title{$O(1)$ benchmarking of precise rotation in a spin-squeezed Bose-Einstein condensate}

\author{Peng Du}
\address{Key Laboratory of Artificial Micro- and Nano-structures of Ministry of Education, and School of Physics and Technology, Wuhan University, Wuhan, Hubei 430072, China}

\author{Hui Tang}
\address{Key Laboratory of Artificial Micro- and Nano-structures of Ministry of Education, and School of Physics and Technology, Wuhan University, Wuhan, Hubei 430072, China}

\author{Jun Zhang}
\address{Key Laboratory of Artificial Micro- and Nano-structures of Ministry of Education, and School of Physics and Technology, Wuhan University, Wuhan, Hubei 430072, China}

\author{Wenxian Zhang}
\email{Corresponding author: wxzhang@whu.edu.cn}
\address{Key Laboratory of Artificial Micro- and Nano-structures of Ministry of Education, and School of Physics and Technology, Wuhan University, Wuhan, Hubei 430072, China}
\affiliation{Wuhan Institute of Quantum Technology, Wuhan, Hubei 430206, China}

\date{\today}

\begin{abstract}
Benchmarking a high-precision quantum operation is a big challenge for many quantum systems in the presence of various noises as well as control errors. Here we propose an $O(1)$ benchmarking of a dynamically corrected rotation by taking the quantum advantage of a squeezed spin state in a spin-1 Bose-Einstein condensate. Our analytical and numerical results show that tiny rotation infidelity, defined by $1-F$ with $F$ the rotation fidelity, can be calibrated in the order of $1/N^2$ by only several measurements of the rotation error for $N$ atoms in an optimally squeezed spin state. Such an $O(1)$ benchmarking is possible not only in a spin-1 BEC but also in other many-spin or many-qubit systems if a squeezed or entangled state is available.
\end{abstract}

\maketitle

\section{Introduction}

High-precision quantum operations are among the most important building blocks for practical quantum computing and quantum information processing~\cite{Nielsen2002}, as well as for entanglement-enhanced quantum sensing beyond the standard quantum limit~\cite{Giovannetti2004Quantum,Xu2019,Ma201189,Bao2020,LM.Duan2000,Pu.H2000}. Characterizing the precision of such quantum operations remains challenging in various physical systems, such as trapped ions, Nitrogen-vacancy centers in diamond, quantum dots in semiconductor, superconducting quantum interference devices, Rydberg atoms in optical tweezers, and ultracold atomic gases~\cite{Gaebler2012,Erika2016,Fogarty2015,Levine2019,Nemirovsky2021}. More efficient and reliable calibrations and benchmarkings of a precise quantum operation are still in demand, particularly for many-qubit or many-particle systems.

A naive method to measure the precision of a quantum operation, e.g., a quantum gate which is described by the gate fidelity~\cite{BOWDREY2002258,Wang2008,NIELSEN2002249}, is to repeat the operation $N$ times and then calculate the fidelity through quantum process tomography~\cite{WuZ2013}. The standard deviation of the fidelity average generally reduces as $1/\sqrt N$. To benchmark a precise quantum operation with a fidelity of $99.9\%$, a million repetitions are usually needed. This is extremely time- and resource-consuming. Improved methods such as randomized benchmarking and its variants are proposed and experimentally realized recently in superconducting quantum interference device, Nitrogen-vacancy centers, and Rydberg atoms, and so on~\cite{Knill2008,ZhangChengxian2016,Wineland2012}. By performing $N$ consecutive random but carefully designed quantum operations, the standard deviation of the operation fidelity may reduce as $1/N$. For instance, Xu {\it et al.} proved that the gate fidelity is above $99.95\%$ by performing roughly $10^3$ random operations for a single qubit realized in Rydberg/neutral atoms trapped in optical tweezers~\cite{pengxu2018}. With this method, it is demonstrated that at least $N$ repeated quantum operations are demanded in order to confirm that the fidelity is above $1-(1/N)$.

To further reduce the repetition number of a precise quantum operation, we propose in this paper a single precise quantum operation to achieve the fidelity with a standard deviation at the level of $1/N^2$ for $N$ entangled particles. With analytical method and numerical simulations, we illustrate this idea by calibrating a precise dynamically corrected rotation (DCR) in an atomic spin-1 Bose-Einstein condensate (BEC) with $N=10^4$ $^{87}$Rb atoms. The rotation error, for a single collective rotation of $N$ atom spins, scales as $\sim 1/N$ for a squeezed spin (entangled) state without noise. Importantly, the rotation infidelity is proportional to the square of rotation error and thus scales as $1/N^2$. In the presence of typical laboratory noise ($\sim$ 0.1 mG) and control imperfection ($\sim$ 1\%), the rotation infidelity is still in the order of $1/N^2$. This efficient calibration method, with a single operation only, can be straightforwardly extended to other many-qubit systems with an entangled quantum state, besides its immediate applications in an atomic spin-1 BEC which has demonstrated paramount potential in entanglement-enhanced quantum sensing~\cite{Stamper-Kurn2013,Kawaguchia2012,Law1998,S.Yi2002,Hamley2012}.

\section{Dynamically corrected rotation of a spin-1 BEC}

We consider a precise rotation of the collective spin of a spin-1 $^{87}$Rb BEC~\cite{Xu2017}. The initial state is an optimal squeezed spin state (SSS) with the optimal squeezing direction along $z$-axis and the mean spin direction on $-y$-axis. We adopt the total atom number $N = 10,000$~\cite{Luo2017}. A strong bias magnetic field $B_0 = 1$ G along $z$-axis is applied to suppress the stay magnetic field noise in the $x$-$y$ plane and to provide a well-defined quantization axis of the system. For such a strong bias field, the second order Zeeman effect, which is about $72$ Hz at $1$ G, must be well controlled. In fact, an additional microwave field is usually employed to cancel the second order Zeeman term in precise spin rotation. To rotate the condensate spin, a driving field $B_x(t)$, which oscillates at a frequency close to the Larmor frequency, is applied along $x$-axis for a certain period $\tau$. As showed in Fig.~1(a) and (b), the condensate spin of the SSS rotates an angle of $\pi$ around $x$-axis, and the mean spin changes from $-y$ to $y$ direction.

The time-dependent Hamiltonian governing the above rotation is
$$H(t) = c_2 {\bf J}^2 + \omega_0 J_z + \Omega_x\cos(\omega t)J_x$$
where $c_2<0$ is the ferromagnetic spin exchange coupling strength, $ \omega_0 = \gamma B_0 = 2\pi \times 0.7$ MHz at $B_0 = 1$ G is the Larmor frequency with $\gamma$ the gyromagnetic ratio of a $^{87}$Rb atom, $\Omega_x = \omega/40$ is the Rabi frequency of a RF field coupled to the condensate atoms, and $J_\alpha=\sum_{i=1}^{N}s_{i\alpha}$ ($\alpha=x, y, z$) is the collective spin of the condensate with $s_{i\alpha}$ the atomic spin-1 matrix for the {\it i}-th atom and $N$ the total atom number. We have set $\hbar = 1$. To obtain this many body Hamiltonian, we have adopted the single-mode approximation which assumes three spin components share the same spatial wave function~\cite{Law1998, S.Yi2002}. We have also neglected the weak magnetic dipole-dipole interaction between atoms since the corresponding time scale is far longer than a spin rotation time $\tau \sim 2\pi/\Omega_x$ (or the trapping potential is spherical). Here we employ the one-axis rotation, instead of the two-axis one, in order to simplify the experimental apparatus and avoid the fine tuning of these driving fields simultaneously perpendicular to each other and to the bias magnetic field.

Adopting the one-axis rotation has two important consequences. One is that the effective Rabi frequency is halved thus $\pi$ pulse duration doubled, due to the relation $\Omega_x \cos(\omega t) e^{-i\omega t} \approx (\Omega_x/2)$ if we drop the fast oscillating term with a frequency $2\omega$. The other is the introduction of the Bloch-Siegert shift, which takes into account of the second order correction of the fast oscillation term and effectively reduces the resonant frequency from $\omega = \omega_0$ to $\omega=\omega_0 - \Delta_{BS}$ with $\Delta_{BS} \approx \Omega_x^2/(16\omega) $~\cite{BlochSiegert1940}. Although it is usually ignored in many driving two-level quantum systems, the $\Delta_{BS}$ must be explicitly included here because of the required high precision of the spin rotation. Given ${\Omega_x}/{\omega}=1/40$, it is easy to check that the relative error of rotation direction is roughly ${\Omega_x}/(16\omega) \approx 0.16\%$ which is already larger than the rotation accuracy $0.1\%$. After taking into account of these two effects, the on-resonance Hamiltonian becomes $H_e = (\Omega_x/2)J_x$ in a rotating reference frame defined by $U_R=\exp[-it(c_2 {\bf J}^2 + \omega J_z)]$. A $\pi$ pulse is realized if $\Omega_x t = 2 \pi$, i.e., $t=\tau$.

\begin{figure}
  \includegraphics[width=3.25in]{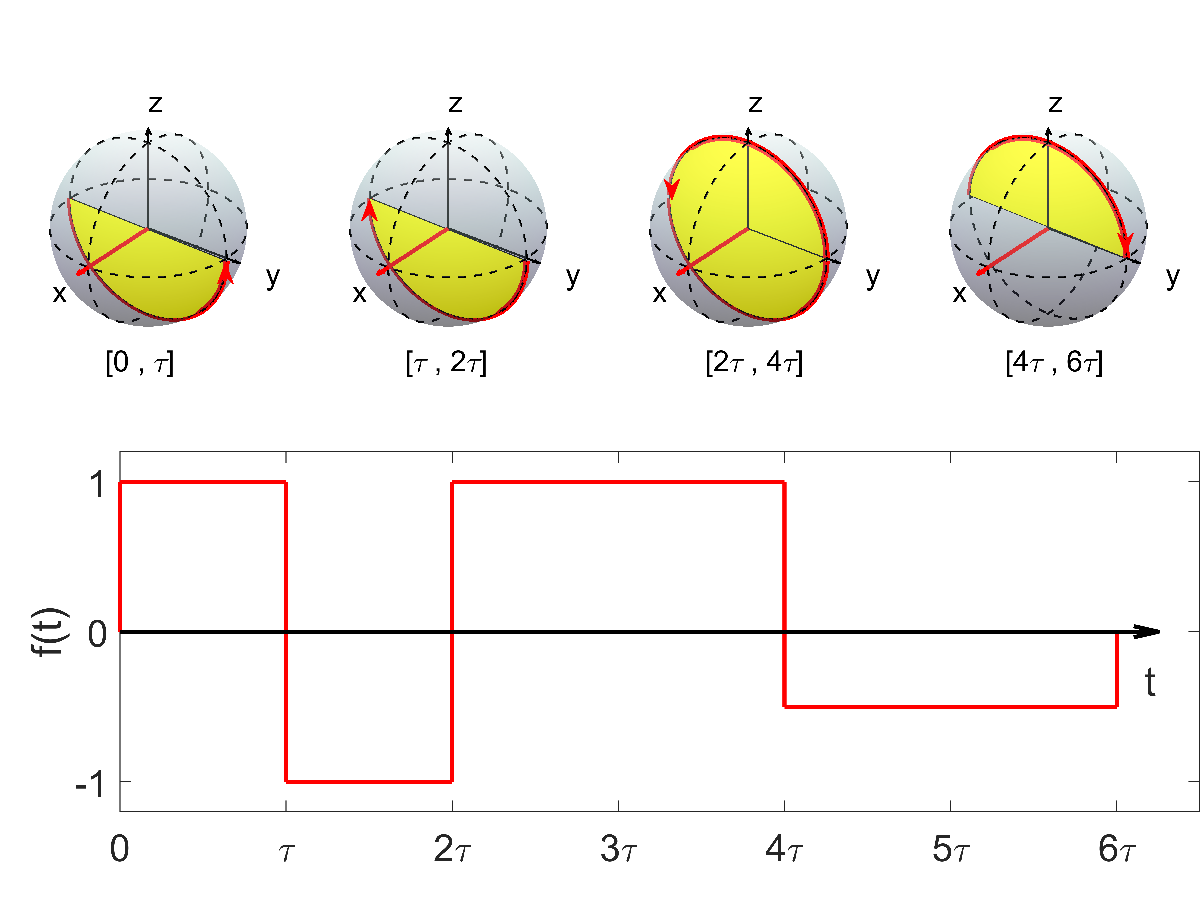}
  \caption{Sequence of DCR. The driving field is $\Omega_x(t) = \Omega_x f(t) \cos(\omega t)$, where the signs of $f(t)$ denote a clockwise and counterclockwise rotation, respectively. The spin is rotated by $\pi$ in the periods $[0,\tau]$, $[\tau, 2\tau]$, and $[4\tau, 6\tau]$ ($\tau={2\pi}/{\Omega_x}$) under the rotating wave approximation, but by $2\pi$ in the period $[2\tau, 4\tau]$.}
  \label{fig:dcr}
\end{figure}

Once we consider a real situation in a BEC experiment, the Hamiltonian for our model must include various noise sources in the laboratory and becomes
\begin{eqnarray}
\label{eq:hL}
  H_{n}(t) &=& c_2{\bf J}^2 + \omega_0 J_z +(\Omega_x +\epsilon)\cos(\omega t)J_x \nonumber \\
   & & + \gamma(b_x J_x+b_y J_y +b_z J_z).
\end{eqnarray}
where $\omega = \omega_0-\Delta_{BS}$ to satisfy the resonant condition, $b_{x,y,z}$ are the three components of a stray magnetic field in the laboratory. We also include explicitly the control error $\epsilon$ caused by the fluctuation of the radio-frequency or microwave power and the finite bandwidth of the control field. We note that the magnetic field noise and the control error are modeled as ensemble white noise, which implies that the stray magnetic field $b_{x,y,z}$ and the control error $\epsilon$ are fixed for a single experiment run but distribute randomly and uniformly from run to run, i.e., $b_{x,y,z}\in [-b_c,b_c]$ and $\epsilon \in [-\epsilon_c, \epsilon_c]$  with $b_c$ and $\epsilon_c$ the respective cutoff.

It is straightforward to obtain the effective Hamiltonian ~\cite{magnus/floquet/fer} with the stray magnetic field and the control error in the rotating reference frame defined by $U_R$
\begin{eqnarray}\label{eq:He}
  H_n &\approx & \frac {\Omega_x} 2 J_x + \left[\frac{\epsilon} 2 + \left(\frac{\Omega_x}{4\omega}\right) \left(\gamma b_z + \frac{\Omega_x^2}{32\omega}\right)\right] J_x \nonumber \\
   && + \left[\gamma b_z - \frac{\epsilon} 2 \left(\frac{\Omega_x}{4\omega}\right)\right] J_z,
\end{eqnarray}
under the conditions $\{\gamma b_c/\Omega_x, \epsilon_c/\Omega_x, \Omega_x/\omega \} \ll 1$. We further write down the evolution operator $U_n = \exp{(- i \tau H_n)} \approx $
$ \exp{[-i \pi  (1 + \epsilon / \Omega_x ) J_n ]}$
where $ J_n \approx J_x + 2 \gamma b_z / \Omega_x J_z$.
It is obvious that the relative error of rotation angle is $\theta_\parallel \approx \epsilon / \Omega_x $ and the relative error of rotation direction $\theta_\perp \approx 2 \gamma b_z / \Omega_x$. Clearly, this imperfect naive rotation~(NR) deviates linearly from an ideal $\pi$ pulse, due to the control error $\epsilon$ and the stray magnetic field $b_z$. The rotation error exceeds 1\% if $\epsilon_c/\Omega_x \sim 1\%$ or $\gamma b_c/\Omega_x \sim 1\%$.

To realize a more precise $\pi$ rotation, we adopt the DCR, which is inspired by the dynamically corrected gate originally designed to suppress static noises, e.g., $b_{x,y,z}$~\cite{Khodjasteh2009PRL,Khodjasteh2009PRA,Khodjasteh2010prl,Stephen1994,WestJacob2010,Kestner2013,WangXin2012,Rong2014,ZengJunkai2018}. It is straightforward to prove that the time-dependent control error is canceled as well by the specific DCR pulse sequence shown in Fig.~\ref{fig:dcr}. In fact, the evolution operator for the DCR cycle is  $U_{DCR} \approx \exp{\{- i \pi  [1 + \theta_{\parallel} ]J_n'\}}$ with $\theta_{\parallel} \approx 7 \Omega_x^2 / (256\omega^2 ) + \gamma b_z/(2\omega)-4\pi (\gamma b_z/\Omega_x)^2 $ and $J_n'=J_x - (4 \pi \gamma b_z \epsilon / \Omega_x^2 )J_y + [4 \pi \gamma b_z \epsilon / \Omega_x^2  - \epsilon / (4\omega) ]J_z$. One immediately finds $\theta_{\parallel} $ and $\theta_{\perp}$ are smaller than that in the NR, indicating the DCR is more accurate (see in Append.~A).

To verify the above analytical results from the Magnus expansion theory, we carry out numerical simulations with the time-dependent Hamiltonian Eq.~(\ref{eq:hL}). For the DCR, the driving amplitude $\Omega_x$ (and the corresponding Bloch-Siegert shift $\Delta_{BS}$) becomes time-dependent as shown in Fig.~\ref{fig:dcr}. Since the relative fluctuation $|\Delta J_{\perp} / \langle {\mathbf J}\rangle | = 1/\sqrt N = 1\%$ for a coherent spin state with $N=10^4$, it is impossible to justify the high accuracy of the DCR. We thus employ an SSS whose relative fluctuation is in the order of $1/N=0.01\%$~\cite{Kitagawa1993, Degen.C.L2017, Ma201189}. Initially, we set the average spin $J_0 = \langle {\bf J}\rangle$ along $-y$-axis and the optimal squeezing direction along $z$ direction. Once the NR or DCR pulses are finished, we calculate the observables $\langle J_z \rangle$ and $\Delta J_z = \sqrt{\langle J_z^2\rangle - \langle J_z\rangle^2}$. The rotation error (precision) is measured by the ratio of the two experimental observables to the average spin $J_0$ which should point along $y$-axis after the rotation. $\langle J_z\rangle /J_0$ denotes the deviation of the spin direction from the ideal one, and $\Delta J_z/J_0$ the quantum fluctuation of the spin. We note that the spin fluctuation along $x$ direction, $\Delta J_x$, is very large (in the order of $N$) and not useful in quantum sensing.

\begin{figure}
  \includegraphics[width=3.25in]{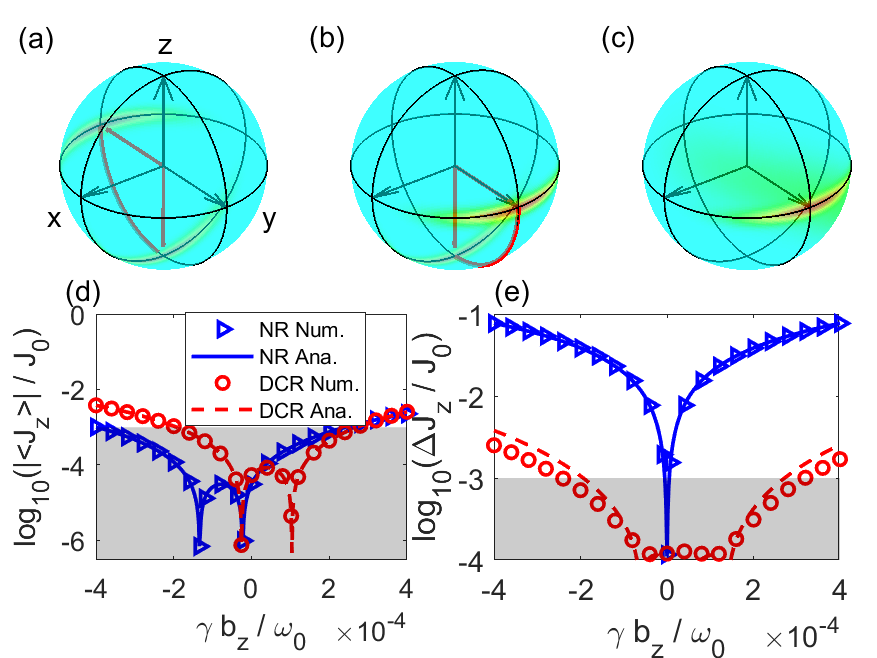}
  \caption{Rotation of the SSS. (a) The initial state with the average spin along $-y$-axis and the optimal squeezing in $z$ direction is rotated around $x$-axis. (b) The SSS is rotated to the final $y$ direction. (c) Comparison of the ideal SSS and the evolved final state with error. (d) Dependence of the average spin direction deviation on the magnetic field noise $b_z$. (e) Spin fluctuation. Panels (d) and (e) show numerical results after NR (blue triangles) and DCR (red circles), and analytical results after NR (blue solid line) and DCR (red dashed line). Other parameters are $N=10^4$, $\omega_0=\gamma B_0$ with $B_0 = 1$ G, $b_{x,y} = 0$, $\epsilon = 0$, and $J_0\approx 0.58N$.}
  \label{fig:pic1new}
\end{figure}

We compare the NR and DCR of the condensate spin at different stray field $b_z$ in Fig.~\ref{fig:pic1new}. We have set $b_{x,y}=0$ and $\epsilon = 0$ for a clear comparison. For the spin average, $\langle J_z \rangle / J_0$, we observe that the deviation from the ideal direction is below 0.1\% if the magnetic field noise is within 0.2 mG, either for the NR or the DCR. For the spin fluctuation $\Delta J_z / J_0$, the minima of both the NR and the DCR are close to the initial value of $1/N$. However, the DCR performs much robust against the field noise than the NR in general. To reach the precision of 0.1\%, the DCR requires the magnetic noise below 0.2 mG but the NR requires much smaller noise. As shown also in Fig.~\ref{fig:pic1new} the numerical results are in good agreement with the analytical ones. It is lengthy but straightforward to obtain the analytical results for $\langle J_z\rangle/J_0$ and $\Delta J_z / J_0$ (Details of the derivation are in Append.~A).

\begin{figure}
  \includegraphics[width=3.25in]{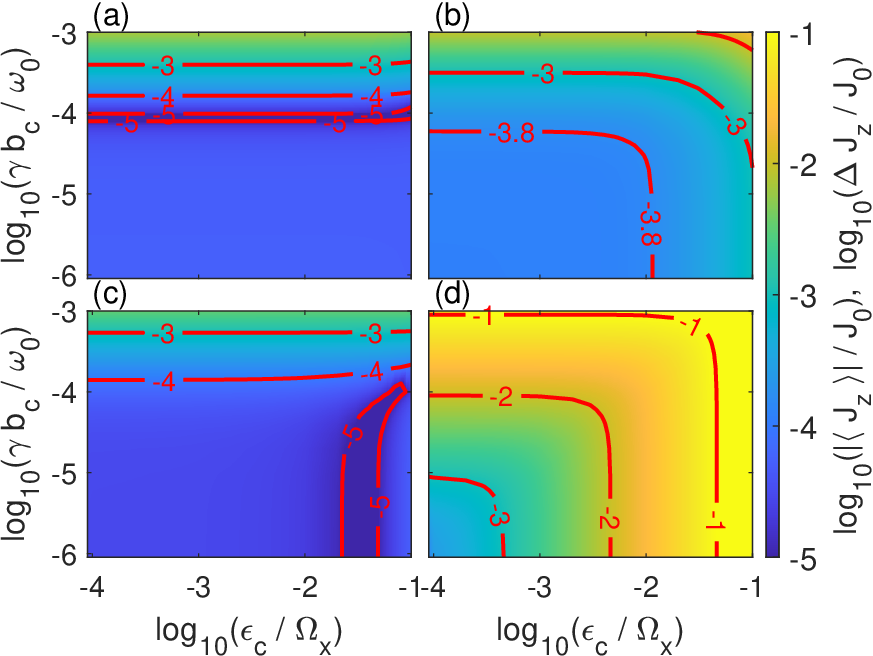}
  \caption{(a) Deviation of the spin direction $\log_{10}(|\langle J_z\rangle|/J_0)$ and (b) spin fluctuation $\log_{10}(\Delta J_z/J_0)$ after the DCR with an initial SSS. (c) and (d) are the same but after the NR. Obviously, the DCR performs more robust against the stray field noise and the control error than the NR, both for the deviation of the spin direction and for the spin fluctuation. Under realistic experiment conditions $b_c \sim 0.1$ mG (i.e., $\gamma b_c /\omega_0 \sim 10^{-4}$) and $\epsilon/\Omega_x \sim 0.01$, the precision of the DCR is well below 0.1\%. Each datum is averaged over $10^6$ random realizations.}
  \label{fig:result}
\end{figure}

In Fig.~\ref{fig:result} we present numerical results of the deviation of the spin direction and the spin fluctuation after the DCR for various cutoff noise strength $b_c$ and control error $\epsilon_c$. For comparison, we also show the same results after the NR. As illustrated in Fig.~\ref{fig:result}(a) and (b), the deviation of the spin direction and the spin fluctuation are small with the stray field noise and the control error. In particular, they are both below 0.1\% if the control error is smaller than $0.01\Omega_x$ and the stray field within 0.1 mG. However, the NR errors shown in Fig~\ref{fig:result}(c) and (d) are rather large, making the NR impossible to estimate the fidelity beyond the standard quantum limit. This is why we adopt the DCR to take the quantum-entanglement advantage of the SSS. We note that the numerical results agree well with the analytical ones which are detailed in Append.~A.

\begin{figure}
  \centering
  \includegraphics[width=3.5in]{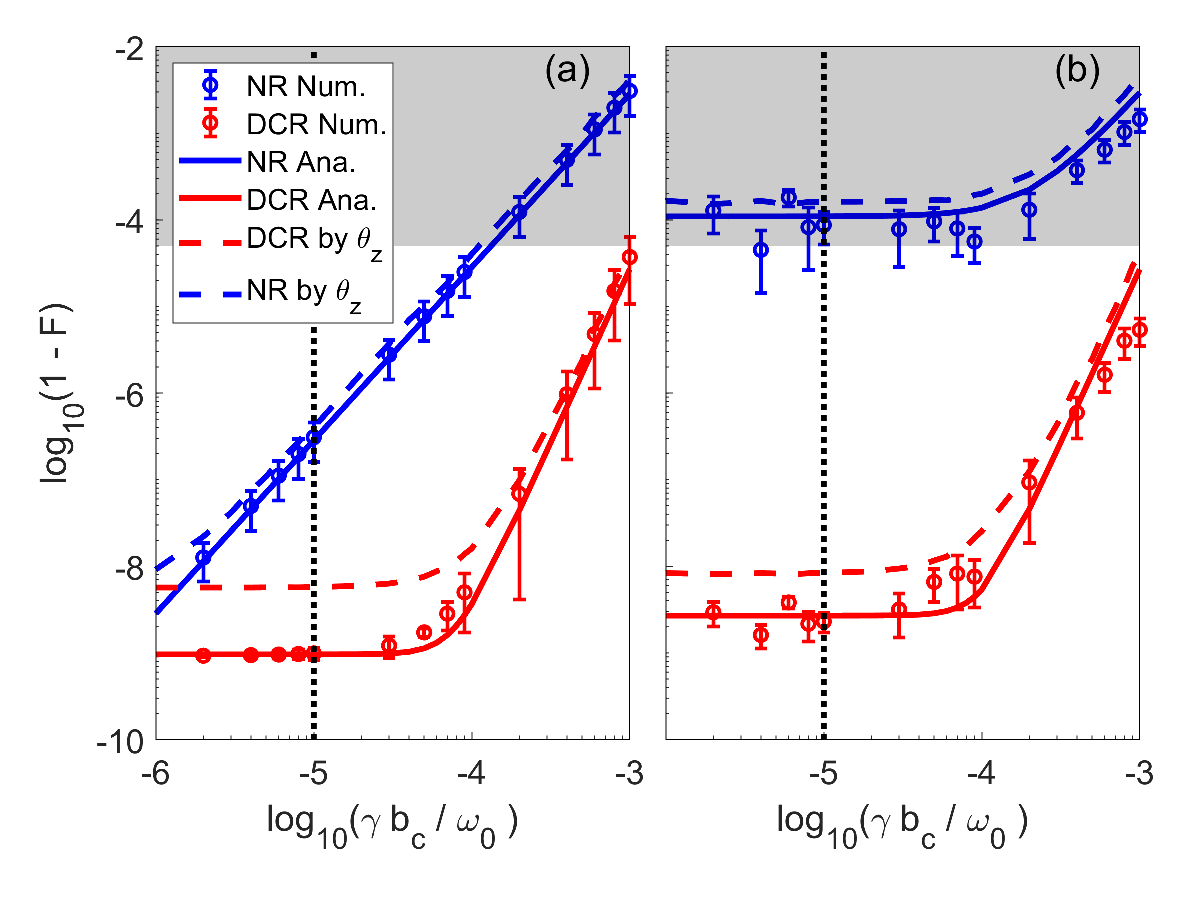}
  \caption{Dependence of the rotation infidelity $1-F$ on the cutoff of the stray magnetic field for (a) $\epsilon_c =0$ and (b) $\epsilon_c/\Omega_x =0.01$. In (a) and (b), the analytical results (solid lines) agree well with the numerical ones (circles with error bars) for DCR (lower) and NR (upper). The dashed lines denote the estimated rotation infidelity by $\theta_z$ for DCR (lower) and NR (upper). Each numerical datum is averaged over $5$ runs of random fields and the standard deviation is denoted by error bars. The vertical and the horizontal boundaries mark the reported $b_c = 0.01$ mG and the infidelity $1-F=5\times 10^{-5}$, respectively~\cite{Rongxing2015}.}
  \label{fig:fid}
\end{figure}

\section{$O(1)$ benchmarking of a precise rotation}

With such a robust and high precision DCR at hand, we compare it with other single-particle quantum operations. The precision of most quantum operations are usually characterized by the operation infidelity, $1-F$, with $F$ the fidelity between the ideal operation and the realized one. After an analytical derivation, we find the rotation infidelity
\begin{equation}\label{eq:err}
  1-F = 1-\left|\frac{\sin{(\frac{3\theta}2)}}{3\sin{\frac{\theta}2}}\right|
\end{equation}
where the fidelity $F=|\Tr(U^\dag X) |/ \sqrt{\Tr(U^\dag U) \Tr(X^\dag X)}$ with $U$ the evolution operator, $X$ the ideal $\pi$-rotation operator, and the rotation error $\theta$ defined by $\exp(-i\theta J_n) = U^\dag X$. The rotation infidelity, $1-F \approx \theta^2/3$ if $\theta \ll 1$. One immediately obtains the rotation infidelity once one knows $\theta$ which may be calculated theoretically, simulated numerically, or estimated (measured) experimentally (More details are in Append.~B).

We present the rotation infidelity after the DCR (and the NR) in Fig.~\ref{fig:fid}. The numerical results are simulated by evolving the system under the time-dependent Hamiltonian Eq.~(\ref{eq:hL}). The analytical ones are straightforwardly calculated with the time-independent effective Hamiltonian Eq.~(\ref{eq:He}) and the application of the DCR pulse sequence. As shown in Fig.~\ref{fig:fid}, the numerical and the analytical results agree well, implying that the effective Hamiltonian is an excellent approximation to the real one if the stray magnetic field and the control error are small. As the stray field decreases from $\sim 1$ mG, the rotation infidelity decreases sharply in the form of $\left( {\gamma b_c}/{\omega_0} \right)^2$ for the NR and $\left( {\gamma b_c}/{\omega_0} \right)^4$ for the DCR. At an extremely small stray field, the infidelity reaches a plateau which stems from the high order terms beyond the Bloch-Siegert shift. Compared to the NR, the rotation infidelity after the DCR is several orders of magnitude smaller if the stray magnetic field lies in the range $b_c \in [0.01, 1]$ mG. In fact, the DCR infidelity at the most stable laboratory field $b_c \sim 0.01$ mG is roughly one thousandth times the reported lowest gate infidelity in Nitrogen-vacancy centers in diamond~\cite{Eto2013,Rongxing2015}, indicating the great potential of the spinor BEC systems in precise quantum operations.

Benchmarking such a small rotation infidelity is a big challenge. However, by noticing the independence of the small rotation error $\theta$ on the atom number $N$ for the DCR, we may make full use of the advantages of many-body entanglement states, e.g. spin squeezed states, to estimate (measure) $\theta$ precisely and calculate the single-particle rotation infidelity $1-F$ with Eq.~(\ref{eq:err}). To estimate efficiently the small rotation error $\theta$, we suggest $\theta \approx \theta_z =  J_z / J_0$ where $J_z$ after the DCR is measured experimentally for an initial SSS with optimal squeezing along $z$-axis. As illustrated in Fig.~\ref{fig:fid}, the rotation infidelity derived from $\theta_z$ agrees well with that from $\theta$, except the region with extremely tiny rotation infidelity $1-F < 10^{-8}$. Such a limitation originates from the spin squeezing limit of the quantum state, which is in the order of $1 / ( 3N^2 ) \sim 10^{-8}$ for $N=10^4$~\cite{Kitagawa1993}. This limitation may be lower than $10^{-8}$ by increasing the atom number in a spin-1 BEC.

We remark that only several measurements of the rotation error are enough to obtain a pretty good estimation of the rotation infidelity with the spin-squeezed quantum state, as shown in Fig.~\ref{fig:fid}. This $O(1)$ measurement requirement greatly relieves the experimental efforts and contrasts sharply to the quantum process tomography and the randomized benchmarking, which require $O(1/(1-F)^2)$ and $O(1/(1-F))$ (equivalent) measurements, respectively~\cite{Rongxing2015,WuZ2013}. Such a huge benefit comes from the accurate estimation of the rotation error and thus the rotation infidelity with the spin-squeezed quantum state, manifesting the quantum supremacy of entangled quantum states over the separable ones. Of course, one may carry out more experimental measurements with the spin-squeezed state to benchmark even lower rotation infidelity beyond the spin squeezing limit $1/(3N^2)$.

\section{Conclusion and discussion}

In conclusion, we propose an $O(1)$ benchmarking method for a precise single-spin rotation with the error derived from the precisely measured rotation error by utilizing a squeezed spin state. With analytical calculations and numerical simulations, we show that a DCR decouples almost perfectly a spin-1 BEC from its magnetic noise environment when performing a $\pi$ rotation. The rotation infidelity after a DCR approaches $10^{-8}$ for $10^4$ atoms at the lowest laboratory magnetic field noise of $\sim 0.01$ mG and a relative control error of $\sim 1\%$. For such a high-precision rotation, it is viable to benchmark it by only several measurements with a squeezed spin state in a spin-1 BEC. Although our example focuses on a precise $\pi$ rotation, the $O(1)$ benchmarking is in principle applicable to an arbitrary rotation with a squeezed spin state, which may be prepared in a spin-1 BEC or a many-qubit system like trapped ions, superconducting qubits, Nitrogen-vacancy centers and neutral atoms in optical tweezers~\cite{Kitagawa1993PRA,Caves1981,Wineland1992PRA}.

The preparation of a spin squeezed state under current experimental condition has been discussed theoretically~\cite{Xu2017,Huang2021}. We notice two recent advances in spinor BECs. (i) Zou {\it et al.} demonstrated spin squeezing in $10^4$ atoms 18~dB below the standard quantum limit (SQL) which is the limit for a coherent spin state~\cite{Zou2018}. (ii) Single-atom level counting was reported via a combination of Stern-Gerlach separation and fluorescence imaging~\cite{Bertrand2021}. By adopting both techniques, it is possible to implement the $O(1)$ benchmarking in spinor BEC experiments, at least at the level of $1-F \sim 10^{-6}$ since the spin squeezing has not reached perfectly the Heisenberg limit (which is 40~dB below the SQL). In addition, entanglement states were generated and high fidelity rotations realized in Nitrogen-vacancy centers experiments~\cite{Xu2017, Rongxing2015,TianyuXie2021,Liu2011,Yang2012pra}, indicating that our method may also be applicable in these many-qubit systems in principle, though the size of qubits is quite limited.

\begin{acknowledgments}

This work is supported by the NSAF (grant No. U1930201), the National Natural Science Foundation of China (grants No. 12274331 and No. 91836101) and Innovation Program for Quantum Science and Technology (grant No. 2021ZD0302100). The numerical calculations in this paper have been partially done on the supercomputing system in the Supercomputing Center of Wuhan University.

\end{acknowledgments}

\begin{appendix}

\section{The Average Hamiltonian}

\begin{figure}
	\includegraphics[width=3.25in]{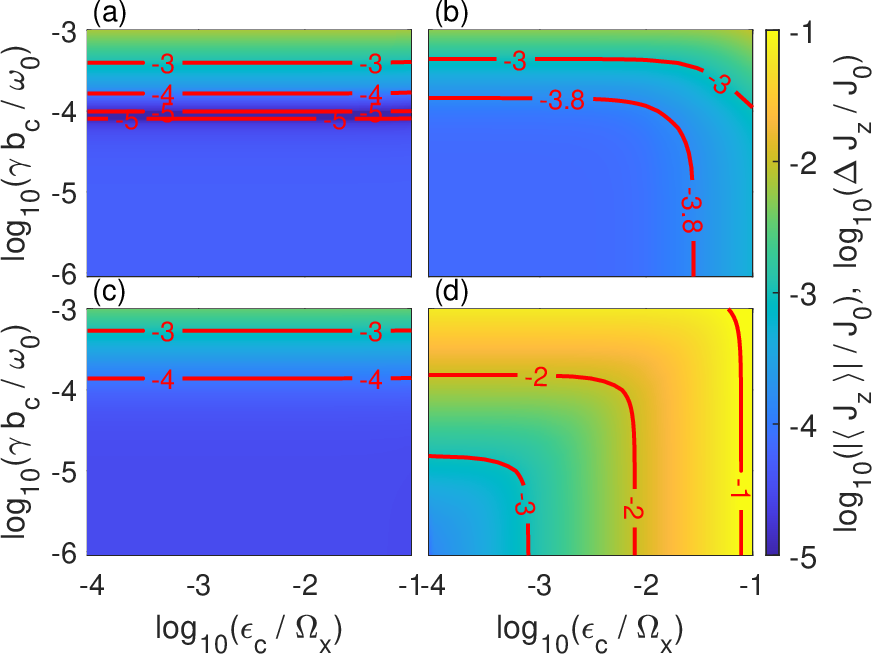}
	\caption{Same as Fig.~3 except for analytical results.}
	\label{fig:anajz}
\end{figure}
The spin-1 Bose-Einstein condensate (BEC) Hamiltonian under the single mode approximation is

\begin{eqnarray}
	\label{eq.hL}
	H_n(t) &=& c_2{\bf J}^2 + \omega_0 J_z +(\Omega_x +\epsilon)\cos(\omega t)J_x \nonumber\\
	&&+ \gamma(b_x J_x+b_y J_y +b_z J_z),
\end{eqnarray}
where $c_2<0$ is the ferromagnetic spin exchange coupling strength, $ \omega_0 = \gamma B_0$ is the Larmor frequency in a magnetic field $B_0$ with $\gamma$ the gyromagnetic ratio, $\Omega_x$ is the Rabi frequency of a RF field (carrier frequency $\omega$) coupled to the condensate atoms, and $J_\alpha=\sum_{i=1}^{N}s_{i\alpha}$ ($\alpha=x, y, z$) is the collective spin of the condensate with $s_{i\alpha}$ the atomic spin-1 matrix for the {\it i}-th atom and $N$ the total atom number. The control error $\epsilon$ and the magnetic field noise $b_{x,y,z}$ are fixed for a single experiment run, but they change for the next. In a typical spin-1 BEC experiment, $\{\gamma b_{x,y,z}/\Omega_x, \epsilon/\Omega_x, \Omega_x/\omega \} \ll 1$.

By employing a rotating reference frame defined by $U_R = \exp{\{- i t (\omega J _z + c_2 {\bf J}^2)}$, the Hamiltonian becomes
\begin{eqnarray}\label{eq.h0(t)}
	H_R(t) &=& \gamma b_z J_z + [(\Omega_x  + \epsilon)\cos(\omega t) + \gamma b_x]\nonumber\\
	&&\times[\cos(\omega t) J_x + \sin(\omega t) J_y]\nonumber\\
	&&+\gamma b_y [-\sin(\omega t) J_x + \cos(\omega t) J_y].
\end{eqnarray}
The average Hamiltonian with Magnus expansion to the second order in a period of $\tau_0 =  2\pi/\omega$ is~\cite{magnus/floquet/fer}
\begin{eqnarray}\label{eq.Heff1}
	\overline H_n &\approx & \left( \frac{\Omega_x}{2} + \frac{\epsilon}{2} + \frac{\gamma b_z \Omega_x}{ 4\omega} + \frac{\Omega_x^3}{128\omega^2} \right) J_x \nonumber\\
	 &&+  (\gamma b_z - \frac{\epsilon \Omega_x}{8\omega}) J_z,
\end{eqnarray}
which is nothing but Eq.~(2). Note that the Bloch-Siegert shift has been included and the third and higher order terms are neglected.

Next we consider the effective evolution operator for the pulse sequence, dynamically corrected rotation (DCR). The effective Hamiltonians $\overline H_k$ during $[k-1,k]\tau$ $(k = 1,2,\cdots, 6)$ are
\begin{eqnarray}
	\overline H_1 &=& \overline H_n, \nonumber \\
	\overline H_2 &\approx & -\left( \frac{\Omega_x}{2} + \frac{\epsilon}{2}  + \frac{\gamma b_z \Omega_x}{ 4\omega} + \frac{\Omega_x^3}{128\omega^2} \right) J_x \nonumber\\
	&& +  (\gamma b_z - \frac{\epsilon \Omega_x}{8\omega}) J_z, \nonumber\\
	\overline H_{3,4} & = & \overline H_n, \nonumber \\
	\overline H_{5,6}  &\approx & -\left( \frac{\Omega_x}{4} + \frac{\epsilon}{2} + \frac{\gamma b_z \Omega_x}{ 8\omega} + \frac{\Omega_x^3}{1024\omega^2} \right) J_x \nonumber\\
	&& +  (\gamma b_z - \frac{\epsilon \Omega_x}{16\omega}) J_z.
\end{eqnarray}
It is straightforward to calculate the effective evolution operator for the DCR
\begin{eqnarray}
	\label{eq.udcg}
	U_{DCR} & = &  e^{(-i2\tau \overline H_5)}e^{(-i2\tau \overline H_1)}e^{(-i\tau \overline H_2)}e^{(-i\tau \overline H_1)}\\\nonumber
	& \approx & e^{\left[-i\pi(1+\theta_\parallel) J_{\bf n}\right]},
\end{eqnarray}
with
\begin{eqnarray}
	\theta_\parallel &=&\frac{7 \Omega_x^2}{256\omega^2}+\frac{\gamma b_z }{2\omega}-\frac{4 \pi(\gamma b_z)^2}{\Omega_x^2},\nonumber\\
	J_{\bf n} & = & J_x - \frac{4 \pi \gamma b_z \epsilon }{ \Omega_x^2} J_y +\left( \frac{4 \pi \gamma b_z \epsilon }{ \Omega_x^2}  - \frac{\epsilon }{4\omega} \right)J_z.\nonumber
\end{eqnarray}
Again we have neglected the third and higher order terms.

With the evolution operator above, one can straightforwardly calculate experimental observables, such as the spin average $\langle {J_{i}} \rangle = \text{Tr} \left( \rho_0 U^\dag {J_i} U \right)$ and its fluctuation $ \Delta {J_{i}} =\sqrt{ \text{Tr} \left( \rho_0 U^\dag {J_i^2} U \right)  - \langle {J_{i}} \rangle^2 }$ with $ i = x, y, z$ for an initial spin state $ \rho_0$. The results with $\epsilon = 0$ are shown in Fig.~2, which agree well with the numerical ones. The results with $\epsilon \neq 0$ are shown in Fig.~\ref{fig:anajz}, which are also close to the numerical ones shown in Fig.~3.

\section{Rotation Infidelity}

\begin{figure}
	\includegraphics[width=0.8 \linewidth]{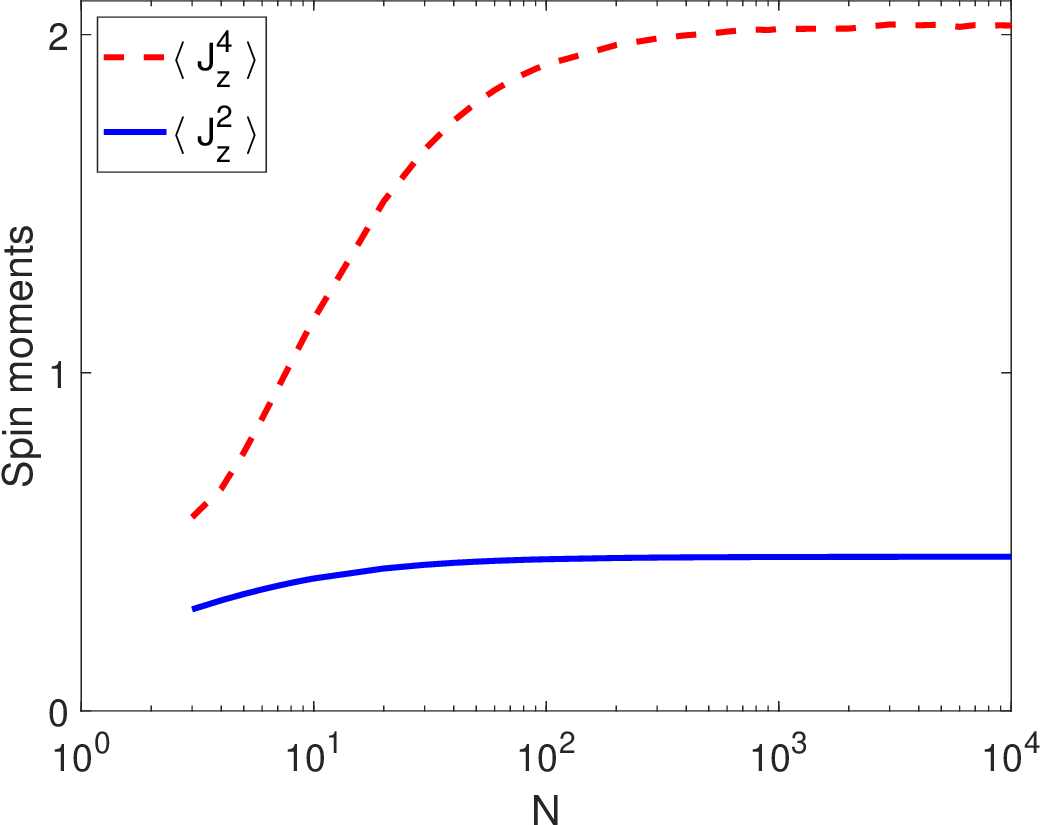}
	\caption{The minimum spin moments of $\langle J_z^2 \rangle$ and $\langle J_z^4 \rangle$ as a function of particle number $N$ for an optimal two-axis spin squeezed state.}
	\label{fig:squeezed}
\end{figure}
Same as a quantum gate fidelity~\cite{Wang2008}, the rotation fidelity for a spin-1 BEC is defined as
\begin{eqnarray}\label{eq.fid}
	F_N & = &\frac{\left|\text{Tr} \left(U^\dag X\right )\right|}{2N+1} = \left|\frac{\sin \left(\left(2N+1\right) \frac{\theta}{2}\right)}{(2N+1) \sin\left( \frac{\theta}{2}\right)} \right|
\end{eqnarray}
where $U = U_{NR} = \exp{(-i\tau \overline{H}_n )}$ for the NR, $U= U_{DCR}$ for the DCR, and $X = \exp{({-i \pi J_x})}$ for an ideal $\pi$ rotation operator around the $x$-axis. We have used that the product of two rotation operators is also a rotation operator, i.e., $U^{\dag}X = \exp(-i\theta J_n)$ with $J_n$ and $\theta$ given by
\begin{eqnarray*}
	J_n &=& \frac{1}{\sin(\theta/2)} \left(\cos{\frac \phi 2} J_x - c \sin{\frac \phi 2} J_y + b \sin{\frac \phi 2} J_z\right), \\
	\cos\left(\frac{\theta} 2\right) &=& a \sin\left(\frac{\phi} 2\right),
\end{eqnarray*}
for a rotation operator $U = \exp(-i\phi J_{\phi})$ and $J_{\phi} = a J_x +b J_y + c J_z$ with $a^2 + b^2 + c^2=1$.

To fairly compare with single atom's rotation fidelity in other systems, it is easy to derive the rotation fidelity for a single spin-1 atom from Eq.~(\ref{eq.fid}), by setting $N=1$,
\begin{eqnarray}\label{eq.fid1}
	F = \frac{2\cos{\theta} + 1}{3} \approx 1-\frac{\theta^2}{3}
\end{eqnarray}
if $\theta$ is small. From the above function, one immediately obtains the average and the fluctuation of the rotation fidelity
\begin{eqnarray}
	F_{\rm avg} &=& \overline{1-\theta^2/3} = 1-\frac{(\Delta\theta)^2 + (\bar\theta\,)^2}{3}, \\
	(\Delta F)^2 &=& \frac 1 9 \left[\overline{\theta^4} -(\overline{\theta^2}\;)^2\right] \nonumber\\
	&=& \frac 1 9 \left[\overline{\theta^4} -\left((\Delta\theta)^2 + (\bar\theta\,)^2\right)^2\right]
\end{eqnarray}
where $\bar x$ is the average of $x$. To calibrate the infidelity to an accuracy $\varepsilon \ll 1$, we need to guarantee that $1-F_{\rm avg} < \varepsilon$ and $\Delta F < \varepsilon$. As shown in the above equation, $1 - F_{\rm avg} $ and $\Delta F$ depends solely on the small error angle $\theta$ which can be estimated experimentally.

To accurately measure the error angle $\theta$, one may take advantage of the squeezed spin state (SSS) in a spin-1 BEC. A crude estimation is $\theta \approx \theta_z = J_z / J_0$ for the designed initial SSS. Under this approximation, we find
\begin{eqnarray}\label{eq.fbar}
	F_{\rm avg} &=& 1 - \frac{(\Delta J_z)^2 + \langle J_z \rangle ^2}{3J_0^2}, \nonumber \\
	\Delta F &=& \frac {1} {3J_0^2}\; \sqrt{\langle J_z^4\rangle - \left[(\Delta J_z)^2 + \langle J_z\rangle^2 \right]^2}\quad .
\end{eqnarray}
The spin moments above can be measured precisely in experiments. It is well known that, for an optimal SSS with $N$ atoms, $\Delta J_z / J_0 \sim 1/N$ and $\langle J_z \rangle = 0$ with optimal squeezing along $z$-axis~\cite{Kitagawa1993PRA}.

For the optimal SSS we employ, we calculate the moments of the condensate spin $J_z$. As shown in Fig.~\ref{fig:squeezed}, $\langle J_z\rangle =0$, $\langle J_z^2\rangle \approx 0.5$, and $\langle J_z^4\rangle \approx 2$ for $N > 10^2$. From Eq.~(\ref{eq.fbar}) one immediately finds that $1-F_{\rm avg}$ and $\Delta F$ are both in the order of $1/N^2$, indicating that $O(1)$ benchmarking at the level of $1/N^2$ is possible if other noises, such as the stray magnetic field and the control error, are well under control with the DCR.

\begin{figure}
	\includegraphics[width=3.25in]{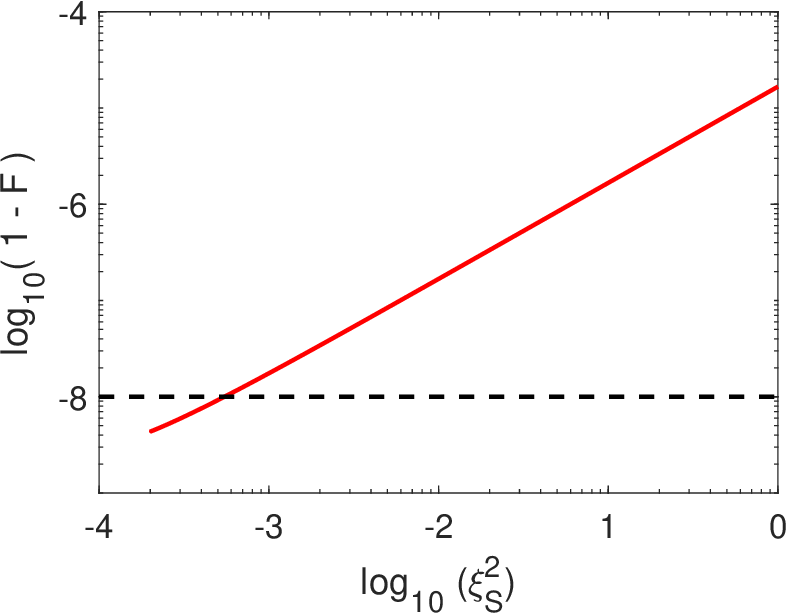}
	\caption{Dependence of infidelity on spin squeezing parameters. The horizontal dashed line denotes the value of $1-F=1/N^2$ with $N=10^4$.}
	\label{fig:fidtosqu}
\end{figure}
\begin{figure}
	\includegraphics[width=3.25in]{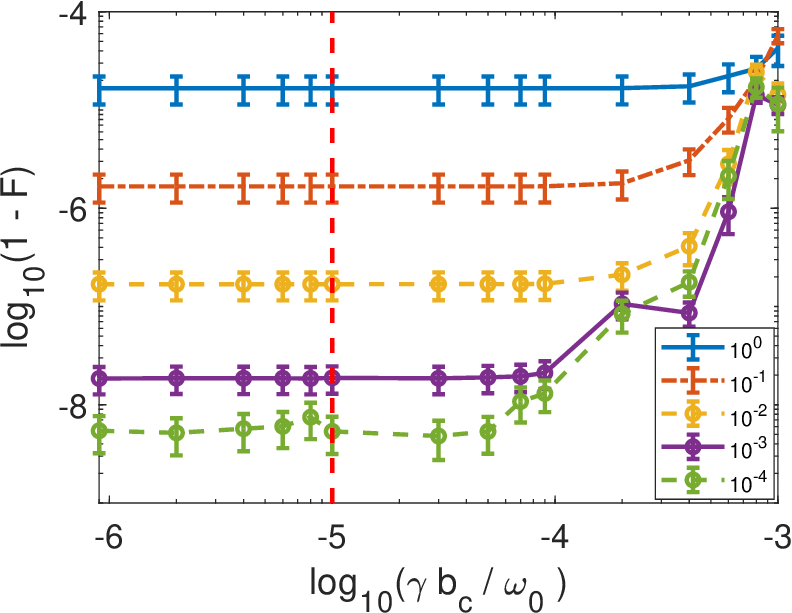}
	\caption{Dependence of infidelity on stray magnetic noises with relative control error $\epsilon_c/\Omega_x = 0.01$ after DCR. The spin squeezing parameter of initial SSS are $1$ (blue solid line), $10^{-1}$ (red dash-dotted line), $10^{-2}$ (yellow dashed line with circles), $10^{-3}$ (purple solid line with circles), $10^{-4}$ (green dashed line with circles), from top to bottom. Each datum is averaged over $5$ random realizations and the error bars are the standard deviation (halved for a clear view). The vertical red dashed line marks the position $b_c = 0.01$ mG.}
	\label{fig:diff_squ}
\end{figure}
In real experiments, initial state often deviates from the optimal SSS. According to Eq.~(\ref{eq.fbar}), the infidelity after a perfect rotation is
\begin{eqnarray}\label{eq.fidxi}
1 - F_{\rm avg} = \frac{N \xi_S^2}{6 J_0^2},
\end{eqnarray}
in the case of zero stray magnetic noises (thus $\langle J_z\rangle = 0$). The spin squeezing parameter $\xi_S^2 = (\Delta J_z)^2/(N/2)$ was originally introduced by Kitagawa and Ueda~\cite{Kitagawa1993}. The result is plotted in Fig.~\ref{fig:fidtosqu}. As shown clearly in the figure, the infidelity reaches $\sim 1/N^2$ when $\xi_S^2$ is smaller than $10^{-3}$. With the presence of magnetic noise and control error, numerical simulations of initial states with various $\xi_S^2$ are shown in Fig.~\ref{fig:diff_squ}. Obviously, the infidelities at small noises agree well with the prediction of Eq.~(\ref{eq.fidxi}), indicating that $O(1)$ benchmarking is possible if the squeezing parameter is smaller than $10^{-3}$.

To estimate $\theta$ more accurately, one may repeat the experiment to measure $\theta_n$ with the optimal squeezing axis of the initial SSS along different directions. In this way, the rotation fidelity becomes more accurate. However, it still scales as $1/N^2$.

To calibrate an arbitrary rotation $R(\alpha, {\bf k})$ with $\alpha$ the rotation angle and ${\bf k}$ the rotation axis, one may prepare the initial SSS under the conditions both ${\bf n}_0 \perp {\bf k} $ and ${\bf s}_0 \perp {\bf k} $ where ${\bf n}_0$ and ${\bf s}_0 $ are the average spin direction and the optimal squeezing direction of the initial state. Here $R(\alpha, {\bf k})$ stands for a $3\times 3$ special orthogonal rotation matrix. Correspondingly, the measurement direction becomes ${\bf s}_f = R(\alpha, {\bf k}) {\bf s}_0$ after the rotation.

\end{appendix}

%\bibliography{newcollection}
%\bibliographystyle{apsrev} % Choose Phys. Rev. style for bibliography, Rev.4

\end{document}